\documentclass[conference]{IEEEtran}

\usepackage{cite}
\usepackage{amsmath,amssymb,amsfonts}
\usepackage{algorithmic}
\usepackage{graphicx}
\usepackage{float}
\usepackage{textcomp}
\usepackage{xspace}
\usepackage{hyperref}
\usepackage{xcolor}
\usepackage{orcidlink}

\usepackage[letterpaper, left=1in, right=1in, bottom=1in, top=0.75in]{geometry}
\usepackage{caption}
\usepackage{balance}

\begin{document}

\title{Multi Kernel Positional Embedding ConvNeXt \\ for Polyp Segmentation}

\author{
\IEEEauthorblockN{Trong-Hieu Nguyen-Mau$^{1,5,6}$\orcidlink{0000-0003-2823-3861}, Quoc-Huy Trinh $^{1,5,6}$\orcidlink{0000-0002-7205-3211}, Nhat-Tan Bui $^{3,5,6}$\orcidlink{0000-0002-4303-1582} \\ Minh-Triet Tran$^{1-6}$\orcidlink{0000-0003-3046-3041}, Hai-Dang Nguyen* $^{2,5,6}$\orcidlink{0000-0003-0888-8908}}
\IEEEauthorblockA{
$^1$\textit{Faculty of Information Technology}\\
$^2$\textit{Software Engineering Laboratory}\\
$^3$\textit{International Training \& Education Center}\\
$^4$\textit{John von Neumann Institute, VNU-HCM} \\
$^5$\textit{University of Science, VNU-HCM}\\
$^6$\textit{Vietnam National University, Ho Chi Minh City, Vietnam} \\
\{20120081,20120013\}@student.hcmus.edu.vn,  1859043@itec.hcmus.edu.vn \\
tmtriet@fit.hcmus.edu.vn, nhdang@selab.hcmus.edu.vn}
}

\maketitle
\begin{abstract}
Medical image segmentation is the technique that helps doctor view and has a precise diagnosis, particularly in Colorectal Cancer. Specifically, with the increase in cases, the diagnosis and identification need to be faster and more accurate for many patients; in endoscopic images, the segmentation task has been vital to helping the doctor identify the position of the polyps or the ache in the system correctly. As a result, many efforts have been made to apply deep learning to automate polyp segmentation, mostly to ameliorate the U-shape structure. However, the simple skip connection scheme in UNet leads to the deficient context information and the semantic gap between feature maps from the encoder and decoder. To deal with this problem, we propose a novel framework composed of ConvNeXt backbone and Multi Kernel Positional Embedding block. Thanks to the suggested module, our method can attain better accuracy and generalization in the polyps segmentation task. Extensive experiments show that our model achieves the Dice coefficient of 0.8818 and the IOU score of 0.8163 on the Kvasir-SEG dataset. Furthermore, on various datasets, we make competitive achievement results with other previous state-of-the-art methods.
\end{abstract}


\section{Introduction}
When cells in the colon or human rectum grow out of control, it can cause electoral cancer, which makes human death \cite{colectoral}. The main reason leading to this disease is Polyps; these objects grow abnormally in the colon and rectum, and over time, these polyps can become cancer. A screening test helps diagnose the polyp to remove it. Moreover, early diagnosis can help some treatment efficiently and prevent the high probability that can cause colon cancer \cite{reason}. One of the most popular screening tests is using an endoscope; this method uses a camera that can capture the situation of the organ in the digestive system in real-time. Therefore doctors can check and diagnose the problem of that digestive system \cite{method_Cancer}.

Several algorithms have been developed in recent years that use Deep Learning (DL) to aid in the early detection of Polyps using endoscopic images. These techniques concentrate on numerous tasks, including classification, segmentation, and detection. Regarding the classification approach, the method that can assist in determining the likelihood of an image having polyps; however, this method cannot visualize the evident site of the symptom, which is why segmentation and detection tasks assist medical professionals in detecting polyps in the digestive system, notably in the colon and the rectum.

UNet and its family structure \cite{unet, unet++, resunet} is a great way to deal with medical segmentation. However, the data on the skip connection has not received enough attention, and many modern approaches are still unable to explore adequate data on the skip connection effectively. Besides, since computer vision has evolved rapidly, many models and modules have been proposed, and one of those is ConvNeXt \cite{convnet}. ConvNeXt achieves state-of-the-art performance on ImageNet, as well as showing great results on COCO detection and ADE20K segmentation while still keeping the simplicity, directness, and efficiency of standard ConvNets.

In this paper, we propose a method that solves the semantic segmentation problem by defining a new way to transfer information in skip connection and utilize the power of ConvNeXt in the segmentation field. Our proposed architecture is the combination of the ConvNeXt \cite{convnet} backbone with the Position Embedding \cite{position-embedding} integrated with multi-kernel-size convolution block, which we call the Multi Kernel Positional Embedding block (MPE), to better supplement the multi-scale context features and the position information for the concatenation features in the skip connection.

\pagebreak

To summarize, our contributions in this paper are threefold: 
\begin{itemize}
    \item We modify the UNet architecture and combine it with ConvNeXt to improve the accuracy of the polyps segmentation task.

    \item We propose the Multi Kernel Positional Embedding block (MPE), which is a better way to utilize the Position Embedding and multi-kernel-size convolution on the architecture to attain impressive results in many metrics.

    \item The effectiveness of our method is demonstrated by extensive experiments and comparisons with previous methods on three different benchmark datasets.
\end{itemize}

The content of this paper is organized as follows. In Section \ref{sec:RelatedWork}, we briefly discuss existing methods for medical segmentation. Then we propose our method in Section \ref{sec:ProposedMethod}. Experiments and results are presented in Section \ref{sec:Experiment}. The ablation study is performed and reported in Section \ref{ablation}. Finally, we conclude our work and suggest new research directions that can be explored for the future in Section \ref{sec:Conclusion}.
\pagenumbering{arabic}
\section{Related Work}
\label{sec:RelatedWork}

An image is segmented by the act of assigning each pixel to a label that has already been proposed. Many studies on the segmentation of various photos with various contents have been done recently \cite{jha2020medico}. The initial techniques for segmenting polyps, such as a fully convolutional network (FCN), trained classifiers to tell one polyp apart from the surrounding area in the image; however, these models have a high error rate. Convolutional neural networks (CNN), with a pre-trained model to recognize and segment polyps, make up most models currently used for segmentation.

UNet is significant to consent that successful training of deep networks requires many thousand annotated training samples \cite{unet}. One crucial modification in our architecture is that in the upsampling part, many feature channels allow the network to propagate context information to higher-resolution layers. Consequently, the expansive path is more or less symmetric to the contracting path and yields a u-shaped architecture. The network does not have any fully connected layers and only uses the valid part of each convolution, i.e., the segmentation map only contains the pixels for which the whole context is available in the input image \cite{unetexplain}.

Recently, many works have been conducted to ameliorate the performance of original U-shape architecture. To replace the original skip connection for reducing the semantic gap between encoder and decoder, UNet++ \cite{unet++} designed a dense connection. Res-UNet \cite{resunet} makes use of residual connections, atrous convolutions, pyramid scene parsing pooling, and multi-tasking inference in addition to a UNet encoder/decoder backbone. A more effective version of Res-UNet architecture for colonoscopic image segmentation is called Res-UNet++ \cite{jha2019resunet++}, which effectively fuses novel modules to improve the encoding and skip operations. A lightweight model with low latency that can be incorporated with low-end endoscope hardware is called Nano-Net \cite{nano}. DDA-Net \cite{dda}, based on a dual decoder attention network, can generate the attention map to improve the semantic representation of the feature maps. However, none of these methods can leverage helpful information of the skip connection.

\section{Method}
\label{sec:ProposedMethod}
Our network is based on UNet with the modification by using the ConvNeXt modules with position embedding and multi-kernel-size convolution. This section summarizes relevant ideas for building an efficient semantic segmentation architecture and describes its key findings.

\begin{figure*}[ht]
    \centering
    \includegraphics[width=1.0\linewidth]{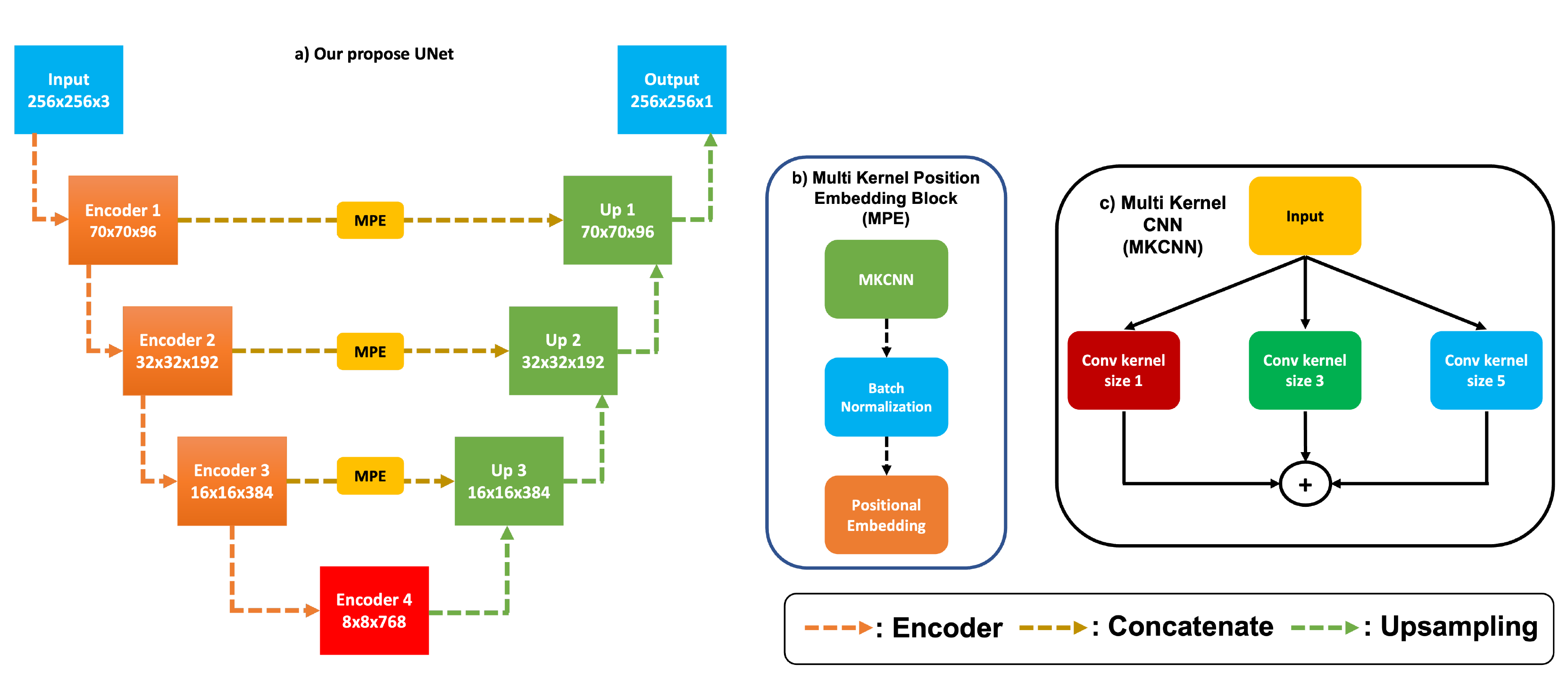}
    \caption{An overview of the proposed method. }
    \label{fig:archi}
\end{figure*}

\subsection{Architecture}

In the original UNet, the Upsampling process follows by the reduction of the channel and increases in the feature size, the feature continuously changes from the $\mathbb{R}^{m\times n \times c}$ to $\mathbb{R}^{m \times 2 \times n \times 2 \times \frac{c}{2}}$ where $m$ is the width of the tensor and $n$ is the height of the tensor with $c$ being the channel at that layer. However, in this paper, we are pleased to create a modified UNet version built with the backbone of ConvNeXt and all the proposed modules, as shown in Fig. \ref{fig:archi}. Foremost, the input comes to the ConvNeXt extractor, and we will get four encoder layers and the last feature map of this backbone. After that, we need to upsample iteratively. We utilize the Convolution Transpose 2D \cite{convtrans} - a better way to upsampling features. Convolution Transpose 2D layer will help the backpropagation process as well as enhance the achieved features.

Before being concatenated with the Upsampling features, encoded features with the shape form $m \times n \times c$ will pass the MPE block. In this block, the features are extracted into a variety of features. The result of the MPE block then is concatenated with the Upsampling as $y_{out} = y_{encoder} + y_{decoder}$ which y is the feature and the underline words are the meaning of each feature. All features continue this step until they go to the last layer of the architecture.

\subsection{ConvNeXt modules for feature extraction}
The authors of ConvNeXt paper \cite{convnet} make some improvements on a standard ResNet \cite{resnet} toward the design of a vision Transformer, and have found several key components that help increase the
performance. This investigation led to the creation of the ConvNeXt family of pure ConvNet models. With a deep
study on macro design, ResNeXt \cite{resnext}, large kernel size, inverted bottleneck, and various layer-wise micro designs, ConvNeXt is an excellent extractor for image classification, detection as well as segmentation. These are the reasons why we choose ConvNeXt as our backbone.

\subsection{Multi Kernel Positional Embedding block}
Multi Kernel Positional Embedding block (MPE), which is also shown in Fig. \ref{fig:archi}, is a module that is combined by two main components: Multi Kernel CNN block (MKCNN) and Positional Embedding.

\vspace{1mm}

The choice of kernel size, pad size, and strides of convolutional layers are critical to the network's ability to capture important oscillatory features. The main idea of the CNN design is the receptive field. We extract meaningful information of encoder features at different scales in this model using multi-kernel-size convolution.

\vspace{1mm}

Multi Kernel CNN block (MKCNN) receives the encoder features; then, this input is extracted using multi-kernel-size convolution. In the end, all the output features from different kernel sizes of convolution layers output sum up all of that to synthesize all the features. The output feature combines different scales of features from Convolution layers and can help the model generalize better information from the encoder features.

\vspace{1mm}

After getting the sum-up output of the MKCNN block, we put it through a BatchNorm layer \cite{batchnorm}. Batch Norm is an important part of modern deep-learning architecture. BatchNorm was recognized as transformational in making deeper neural networks that could be trained faster and more stable.

\vspace{1mm}

Previous works \cite{islam, amirul, semih} have shown the existence of absolute positional information in CNN's latent representations and after the global average pooling (GAP) layers. Islam et al.\cite{amirul} also explore how much absolute position information can be extracted from various classification and segmentation pre-trained CNN models. Based on their experiment, the location classification accuracy using pre-trained image segmentation is almost perfect. Given that the model aims to learn as much positional information as possible, this demonstrates the necessity of absolute positional information for image segmentation tasks. 
Additionally, it is shown that positional information enhances task performance for semantic segmentation and instance segmentation \cite{amirul1}. Inspired by those works, we also use the Positional Embedding layer to enhance the absolute positional information. The Positional Embedding layer is computed through equation \ref{equation1}:
\begin{center}
    \begin{equation}
        PE(x,y) = \sin{\frac{pos}{1000\frac{2i}{d_{model}}}}
        \label{equation1}
    \end{equation}
\end{center}
where $pos$ is the position index and $i$ is the dimension index. 

In our method, we apply Positional Embedding to the Vision task. First, a position embedding mask is created by the computation from the feature of the image. Then it is added to the feature in previous. Therefore, this can help the model get the position information for each part of the feature. 

Putting the positional information into a more meaningful feature could be used to study the data greater. In addition, this computation will help improve the quality of the mask generated by the model.

\subsection{Loss Function}
For the architecture training, we propose to use the Jaccard Loss Function \cite{jaacaard} with the following formula \ref{jac}: 

\begin{equation}
\centering
    JaccardLoss(y,\hat y) = \alpha*(1-\frac{\alpha+\sum_{c}^Cy_{c}*\hat y_{c}}{\alpha +\sum_{c}^Cy_{c} + \hat y^{c} - y_{c}*\hat y_{c}})
\centering 
\label{jac}
\end{equation}

This loss function not only enables the model to learn better but also controls the model's performance on the pitch of the tissues. The Jaccard Loss \cite{jaacaard} is mentioned as the IOU metric, with y as the ground truth and the predicted is $\hat y$; these two labels are demonstrated in the one-hot vector to present and classes $C$ being their length. However, to prevent the vanishing gradient, there is a smoothing factor called alpha $\alpha$, which helps the training result be generalized.

\vspace{3mm}

\section{Experiment}
\label{sec:Experiment}

\subsection{Dataset}
\begin{figure} [H]
    \centering
    \includegraphics[height=2cm]{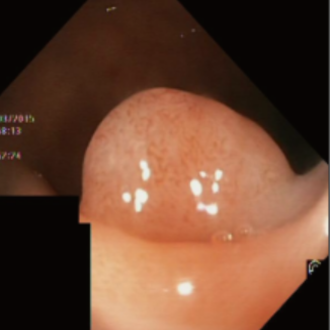}
    \includegraphics[height=2cm]{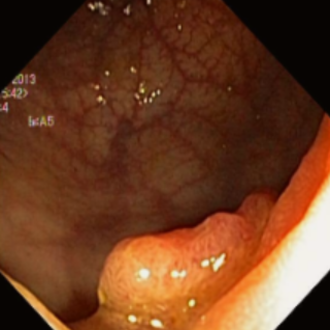}
    \includegraphics[height=2cm]{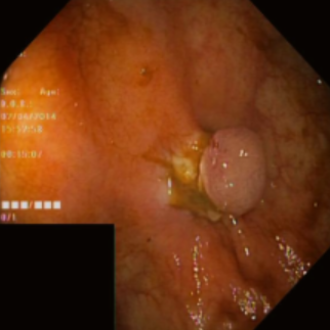}\\
    \includegraphics[height=2cm]{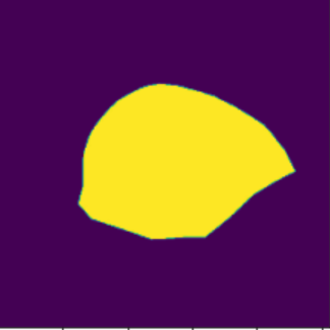}
    \includegraphics[height=2cm]{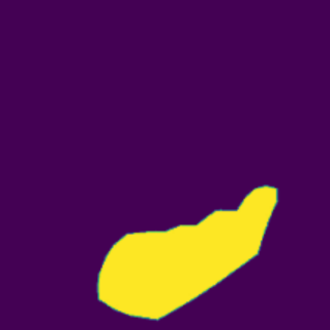}
    \includegraphics[height=2cm]{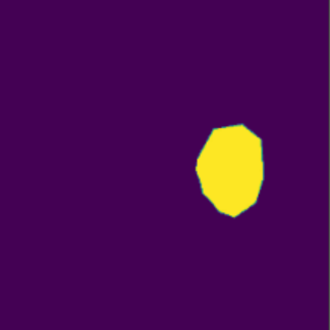}
    \caption{Polyps and their masks from Kvasir-SEG dataset}  
    \label{fig:images}
\end{figure}

The dataset for the experiment is the Kvasir-SEG dataset \cite{kvasir-seg}. This dataset contains 1000 polyp images and their corresponding ground truth from the Kvasir Dataset v2. The Kvasir-SEG dataset appreciates researching and developing new and improved methods for segmenting polyps. For example, Fig. \ref{fig:images} demonstrates some samples of the Kvasir dataset used for the training process.

\vspace{1mm}

For study purposes, we split the Kvasir dataset into training, validating, and testing. We conduct experiments on all these three parts. The training, validating, and testing make up 60\%, 20\%, and 20\%, respectively.


\subsection{Augmentation}
While training the model, we also apply many augmentation techniques to improve the variousness of the dataset. Doing augmentation also increases the data distribution complexity, which has a positive effect on the dataset's domain \cite{augmentation}. To enrich the dataset, we use Center Crop, Random Rotate, Grid Distortion, Cut Out \cite{devries2017improved}, Horizontal, and Vertical Flip to improve the diversity of the dataset.

\begin{figure} [H]
    \centering
    \includegraphics[height=2cm]{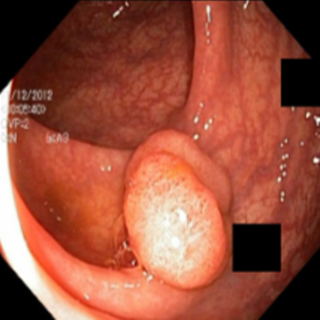}
    \includegraphics[height=2cm]{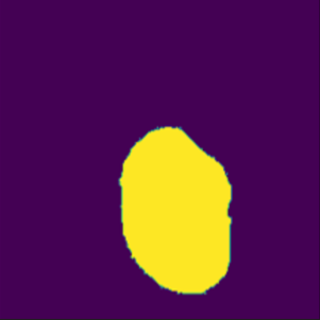}
    \caption{Polyps and their masks after apply Cut Out augmentation}  
    \label{fig:cutout}
\end{figure}

\vspace{1mm}

Fig. \ref{fig:cutout} is the sample when applying Cut Out augmentation \cite{devries2017improved}; this advanced method adds the noise in the area with all pixel values are 0 to the image randomly. Then the mask has also added to that area in the same position.
\subsection{Implementation details}
All architectures and modules are implemented using the Keras framework with the Tensorflow backend. The input images are normalized to [-1, 1]. All models are trained with reported augmentations. We use Adam optimization \cite{kingma2014adam} with an initial learning rate of 0.0001. Then we apply the Cosine Annealing learning rate schedule to prevent the model from getting stuck in a local cost minimum. Our experiment is performed on double NVIDIA Tesla P100 16GB. We use the batch size of 128, and the training time is 5 hours for the entire dataset. Last but not least, our model is tuned by 270 epochs to get the best result.

\vspace{-4mm}
\subsection{Metrics}

IOU and Dice Coefficient metrics are used to evaluate our method's performance. These two metrics evaluate the ground truth with the predicted mask from the test dataset.

The following is the formula of IOU \cite{iou}:
\begin{equation}
    \centering
    IOU = \frac{\textrm{Area  of  Overlap}}{\textrm{Area of Union}}
    \label{iou}
\end{equation}

The formula \ref{iou} demonstrates the $\textrm{Area of Overlap}$ is the common area of two predicted masks, and the $\textrm{Area of Union}$ is all of the areas of two masks.

Regarding the Dice Coefficient \cite{dicecoef}, it has the following formula:
\begin{equation}
    \centering
    Dice Coefficient = \frac{2 * |X \cap Y|}{|X \cup Y|}
    \label{dicecoef}
\end{equation}

The formula \ref{dicecoef} computes the division between the common area of two masks with the union area of two masks where X and Y are for 2 masks.

\vspace{-4mm}
\subsection{Result}
\subsubsection{Results on the Kvasir-SEG dataset}
\begin{figure} [H]
    \centering
    \includegraphics[height=6.3cm]{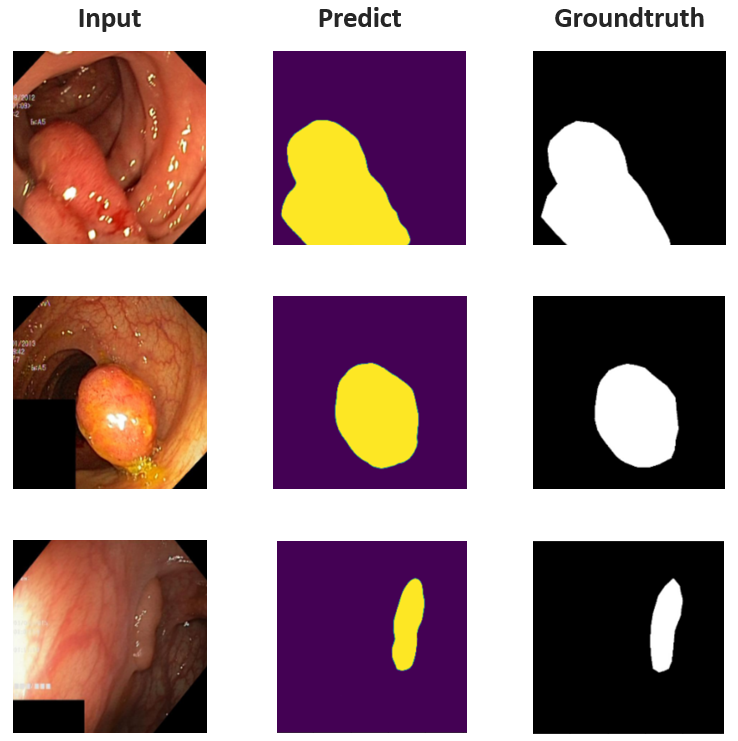}
    \caption{Input image, prediction output, and corresponding groundtruth from Kvasir-SEG}  
    \label{fig:result}
\end{figure}

Our proposed method shows great performance on the test dataset, which is presented in Fig. \ref{fig:result}. We could say that our network produces significant and qualitive segmentation masks on the Kvasir-SEG dataset. It also achieves the IOU score of 0.8163 and the Dice coefficient score of 0.8818.

\subsubsection{Results on the Automatic Polyps dataset}
The Automatic Polyps dataset \cite{jha2020medico} is derived from the Medico Automatic Polyp Segmentation Challenge aimed at developing a computer-aided diagnostic system for automated polyp segmentation to detect all types of polyps. As shown in table \ref{tab:tab2}, our model achieves very good results, comparing Res-UNet \cite{resunet} and NaNo-Net-A \cite{nano} we exceeded the score.
\begin{table}[H]
\begin{center}
\begin{tabular}{| c | c | c |} 
 \hline
 Model & IOU & Dice coef \\ [0.5ex] 
 \hline
 Ours & \textbf{0.8001} & \textbf{0.8630} \\
 \hline
 Res-UNet \cite{resunet} & 0.7396 & 0.8154 \\
 \hline
 Nano-Net-A \cite{nano} & 0.6319 & 0.7364 \\
 [1ex] 
 \hline
\end{tabular}
\end{center}
\caption{Results on Automatic Polyps dataset \cite{jha2020medico}}
\label{tab:tab2}
\end{table}
\subsubsection{Results on the EndoTect 2020 dataset}
The EndoTect 2020 dataset \cite{endotect} is introduced in the EndoTect Challenge. This challenge aims to facilitate the development of algorithms to help medical professionals uncover commonly occurring abnormalities in the gastrointestinal tract. Again, PEFNet performed surprisingly well with his IOU score of 0.7607 and Dice coefficient score of 0.8406, outperforming both DDA-Net\cite{dda} and Nano-Net\cite{nano}. The results are shown in table \ref{tab:tab3}.
\begin{table}[!h]
\begin{center}
\scalebox{1}{%
\begin{tabular}{| c | c | c |} 
 \hline
 Model & IOU & Dice coef \\
 \hline
 Ours & \textbf{0.7967} & \textbf{0.8565} \\ 
  \hline
 DDA-Net \cite{dda} & 0.7010 & 0.7871  \\
  \hline
 Nano-Net \cite{nano} & 0.6471 & 0.7518\\[1ex] 
 \hline
\end{tabular}
}
\end{center}
\caption{Results on EndoTect 2020 dataset \cite{endotect}}
\label{tab:tab3}
\end{table}%
\subsubsection{Results on the Kvasir Sessile dataset}
The Kvasir Sessile dataset \cite{jha2021comprehensive} is published separately as a subset of Kvasir-SEG \cite{kvasir-seg}. This record was selected with the help of an experienced gastroenterologist, which means that the label quality is much better. As shown in table \ref{tab:tab4} In addition, our method achieves the second highest IOU score after DDA-Net \cite{dda} and the highest Dice coefficient score \cite{jha2019resunet++}.
\begin{table}[!h]
\begin{center}
\scalebox{1}{%
\begin{tabular}{| c | c | c |} 
 \hline
 Model & IOU & Dice coef \\ [0.5ex] 
 \hline
 Ours & 0.8322 & \textbf{0.8967} \\
  \hline
 Res-UNet++ \cite{jha2019resunet++} & 0.7927 & 0.8133 \\
  \hline
 DDA-Net \cite{dda} & \textbf{0.8576} & 0.8201 \\
  \hline
 Res-UNet++ +TGA+CRF \cite{jha2019resunet++} & 0.7952 & 0.8508 \\
 [1ex] 
 \hline
\end{tabular}
}
\end{center}
\caption{Results on Kvasir Sessile dataset \cite{jha2021comprehensive}}
\label{tab:tab4}
\end{table}%

 Compete with other models such as Res-UNet \cite{resunet}, DDA-Net \cite{dda}, Nano-Net \cite{nano}, Res-UNet++ \cite{jha2019resunet++}, our networks show great metric results on different datasets. This indicates that the model generalizes well to various datasets and validates the effectiveness of location embedding methods and multi-kernel-size convolution in segmentation tasks.

\section{Ablation Study}
\label{ablation}
After training and evaluating on Kvasir-SEG, we get the results shown in table \ref{tab:tab1}. The traditional method uses the ConvNeXt backbone to achieve the IOU score of 0.7048 and the Dice coefficient of 0.7711. This greatly improved the original UNet \cite{unet} baseline. This shows that ConvNeXt works well as an encoder. Furthermore, after adding the MPE block to the architecture, the IOU score improved to 0.7681 on the test dataset. The larger ConvNeXt version gives significantly better results. Finally, with ConvNeXt-Base as the encoder achieves the IOU score of 0.8163 and a Dice coefficient of 0.8818.

\begin{table}[!h]
\begin{center}
\scalebox{1}{%
\begin{tabular}{|c| c| c| c|} 
 \hline
 Model & Parameter & IOU & Dice coef \\ [0.5ex] 
 \hline
 Baseline & - & 0.4882 & 0.7289 \\ 
 \hline
 ConvNeXt-Tiny & 38M & 0.7048 & 0.7711 \\ 
 \hline
 ConvNeXt-Tiny+MPE & 52M & 0.7681 & 0.8301 \\ 
 \hline
 ConvNeXt-Small+MPE & 74M & 0.7895 & 0.8692 \\
 \hline
 ConvNeXt-Base+MPE & 129M & \textbf{0.8163} & \textbf{0.8818} \\ [1ex] 
 \hline
\end{tabular}
}
\end{center}
\caption{Our Performance Results on Kvasir dataset.}
\label{tab:tab1}
\end{table}

\vspace{-4mm}
\section{Conclusion}
\label{sec:Conclusion}
In brief, we propose the Multi Kernel Positional Embedding block (MPE) combined with the ConvNeXt backbone to deal with the segmentation task. We utilize the multi-kernel size convolution to capture a better receptive field and multi-scale features, together merging them with the positional information, which leads to better results. This network achieves excellent performance on many datasets. With the encoder of ConvNeXt-Base, we gain a 0.8163 IOU score and 0.8818 Dice coefficient on the Kvasir-SEG dataset. This is the evidence that our modifications help to enrich the information in generating the segmentation mask. With this article, we can give more ideas to achieve better performance in polyp segmentation tasks based on the multi-kernel and Positional Embedding technique. The absolute position information can be explored further to develop a better segmentation model.

\section*{Acknowledgement}
This research is funded by University of Science, VNU-HCM, under grant number T2022-84.
\balance
\bibliographystyle{IEEEtran}
\bibliography{myref.bib}

\end{document}